# Stress-sign-tunable Poisson's Ratio in Monolayer Blue Phosphorus Oxide

Bowen Zeng[a], Mengqiu Long [a,b,*], Yulan Dong[a], Jin Xiao[c], Shidong Zhang[a], Yougen Yi[a], and Yongli Gao[a, d]

[a]*Hunan Key laboratory of Super Micro-structure and Ultrafast Process, School of Physics and Electronics, Central South University, Changsha 410083, China*
[b]*Institute of Low-dimensional Quantum Materials and Devices, School of Physical Science and Technology, Xinjiang University, Urumqi, 830046, China*
[c]*School of Science, Hunan University of Technology, Zhuzhou 412007, China*
[d]*Department of Physics and Astronomy, University of Rochester, Rochester, NY 14627, USA*
*E-mail: mqlong@csu.edu.cn

## Abstract

Negative Poisson's ratio (NPR) materials have attracted tremendous interest due to their unusual physical properties and potential applications. Certain two-dimensional (2D) monolayer materials have also been found to exhibit NPR and the corresponding deformation mechanism varies. In this study, we found, based on first-principles calculations, that the Poisson's ratio (PR) sign of monolayer Blue Phosphorus Oxide (BPO) can be tuned by strain: the PR is positive under uniaxial strain $\varepsilon <= -1\%$ but becomes negative under $\varepsilon > 0$. The deformation mechanism for BPO under strain depends on the mutual competition between the P-P attraction and P-O repulsion effect, and these two factors induce two different deformation pathways (one with positive PR, and the other with NPR). Moreover, with increasing of strain, both the decreased strength of P-P attraction and the increased strength of P-O repulsion effect modulate the PR of BPO from positive to negative.

# Introduction

The Poisson's ratio (PR), also called the transverse deformation coefficient, refers to the negative ratio between the transverse strain and longitudinal strain in elastic loading.[1,2] When subjected to strain in the longitudinal direction, the materials with a positive Poisson's ratio (PPR) tend to shrink laterally; in contrast, the materials with a negative Poisson's ratio (NPR) expand laterally. Besides exhibiting a counterintuitive structure deformation under the strain, NPR materials also possess many specific physical properties, such as enhanced shear resistance,[3] indentation hardness[4,5] and fracture toughness,[6] and thus have a wide range of applications. Since Lakes presented a novel re-entrant structure with a NPR in his seminal work,[7] many efforts have been made to develop NPR materials theoretically and experimentally.[8-10]

A variety of NPR materials have been developed, such as three-dimensional bulk material, metal nanoplate, quasi-2D systems and 2D monolayer materials.[11-17] The underlying mechanisms for the NPR behavior are intriguing and should be specifically analyzed. In bulk materials, the corresponding mechanism for explaining the NPR behavior and the principles for developing new NPR materials are based primarily on specific microstructure,[18-22] hence it is natural to think that the magnitude or sign of PR can be modulated by geometry structural transition.[9,20,23] As for metal nanoplate, surface stress effects are also important, which combined with stress-induced phase transitions causes NPR.[24,25] Some 2D monolayer materials have also been found to exhibit NPR behavior and the corresponding deformation mechanism varies. For example, the NPR behaviors in borophene, phosphorene and phosphorene-like monochalcogenides are attributed to the intrinsic crystal pattern.[26-29] Owing to small bending modulus,[30,31] the NPR of 2D materials can be also induced by the structural rippling.[32-34] It should be noted that very few studies have examined the NPR effect on other physical properties in monolayer materials.[35] There are still much need to be done to develop the connection between the NPR and other physical properties theoretically and examine it experimentally for monolayer materials. It have been reported that NPR membranes with pore is superior to conventional membranes in filter defouling.[36] Experimentally, monolayer graphene with nanopore has been fabricated successfully for DNA sequencing.[37] These applications suggest that the monolayer NPR materials with nanopore have potential in smart filtration applications, and we also expect that more applications will emerge in monolayer NPR materials due to their reduced dimensionality.

Recently, Yu *et al* have found an intrinsic in-plane NPR in 1T two-dimensional transition metal dichalcogenides (TMDs) and demonstrated that the corresponding NPR behavior is dominated by the electronic effect rather than merely by their geometric structure.[38] Inspired by their work, we wonder whether there are other electronic effect inducing NPR, and whether this electronic effect can be tuned by strain. It has been reported that edge sulfur passivation can induce NPR in zigzag blue phosphorus (BlueP) nanoribbon, because of the particularity of the P–S bonds at edges,[39] indicating that covalent modification has a great influence on the mechanical properties of BlueP. As we know, BlueP is one of four types of stable allotrope of phosphorus, particularly noteworthy its binding energy is at most 2 meV per atom higher than that of black phosphorus,[40, 41] indicating that BlueP is as stable as black phosphorus. BlueP is a semiconductor with 2.0 eV indirect band gap,[42] while the electronic structure of BlueP can be changed dramatically when covalently modified by oxygen; BPO has been proved to be a semiconductor with direct band gap.[43] The synthesis of BlueP require a substrate,[44] like the preparation of two-dimensional silicene[45] and germanene,[46] while the covalent functionalization of germanene can be synthesized in gram-scale quantities and exfoliated into single layers,[47] suggesting that it may be easier to synthesize the BPO than BlueP due to the enhanced stability.

The specific feature of BPO is not limited to the band structure transition. In this study, we found that the PR of BPO is positive under uniaxial strain $\varepsilon <= -1\%$ but becomes negative under $\varepsilon > 0\%$, suggesting BPO would be expanded laterally regardless of longitudinal compression (strictly speaking, the compressive strain need to be greater than -1% ) or extension. The structural relaxation under strain depends on the mutual competition between P-P attraction and P-O repulsion effect, and the different response of their strength to strain cause the sign change of PR.

## Calculation methods

Our calculations are performed with density functional theory (DFT) via the Vienna ab initio simulation package (VASP).[48] We adopt the generalized gradient approximation (GGA)[49] with the Perdew–Burke–Ernzerhof (PBE) exchange-correlation potential. The projector-augmented wave (PAW)[50] is used with a cut-off energy of 600 eV. The criterion of convergence for structure relaxation is that the change of total energy is less than $10^{-7}$ eV and the residual force on the atom less than 0.001 eV $Å^{-1}$. The Brillouin zone (BZ) is sampled by $47\times47\times1$ and $47\times31\times1$ for structural relaxation for unit cell and rectangle supercell respectively. The length in the out-of-plane direction is set as 20 Å to ensure a large vacuum layer. The charge density is plotted by Visualization for Electronic and Structural Analysis (VESTA).[51] The elastic constant are calculated via VASP[48] with parameter IBRION = 6. In this method, the elastic constant can be formally decomposed into a purely electronic term with frozen ion and into a ion-relaxation term due to internal atomic displacements upon strain. The former term is defined as the second derivative of the total energy versus the lattice vector variation with fixed atomic coordinates, while the latter term are determined by inverting the ionic Hessian matrix and multiplying with the internal strain tensor.[52, 53]

## Results and discussions

The atomic structure of BPO is depicted schematically in Fig. 1(a, b). The optimized lattice constant of BPO is 3.669 Å with hexagonal lattice, which is larger than lattice constant of BlueP (3.33 Å),[40] in good agreement with previous theoretical study.[43] The lattice vector *a* is set parallel with x direction (i.e. zigzag chain) and *b* parallel with y direction (i.e. armchair chain). In pristine BlueP, each P atom has five valence electrons, three of them covalently bonding with three near adjacent P atoms, leaving two electrons with high chemical activity. In BPO, this lone pair is bonded to oxygen (P=O bond). To better understand the form of P=O bond and other bonds in BPO, we have plotted the corresponding band-decomposed charge density (modulus squared of Bloch wavefunction) of all occupied state (Fig. S1), and choose three typical charge density distribution which can be analogized to molecular orbital as shown in Fig. 1(c, d, e). It was observed that 1-st band is dominated by $p_z$ and $s$ orbital of P and O atom, and its electron density is mainly localized between P and O atom, namely P-O bonding state. The 11-th band is dominated by $p_{x(y)}$ of O atoms and $p_{x(y)}$ of P atoms, and the bonding state is formed between P-P while the antibonding state is formed between P-O. Being the lone pair of O atom, the 8-th band's charge density exhibits obvious nonbonding state.

Although the periodic structure is distinct from the molecule, the binding state (bonding state or antibonding state) between atoms should still follow these two rules: the bonding state (antibonding state) make the atoms attract (repel) each other and the strength of bonding state or antibonding state is inversely proportional to the distance between the atoms. Combining charge density distribution of all occupied states, we can draw some important conclusions: (i) The P-P bonding state cause P-P attraction. (ii) Unexpectedly, there are some antibonding states between P-O, which show a complex instead of double bond between P-O. (iii) Other than the P-O bonding state, it was found that the charge density distribution of occupied state shows a distinct separation between P-P plane and O (Fig. S1), which also can be considered as, to some degree, a certain antibonding state between P-P plane and O. These antibonding states bring the P-O repulsion effect, which will be confirmed by the simulation in supporting materials (Fig. S2) and the following analysis.

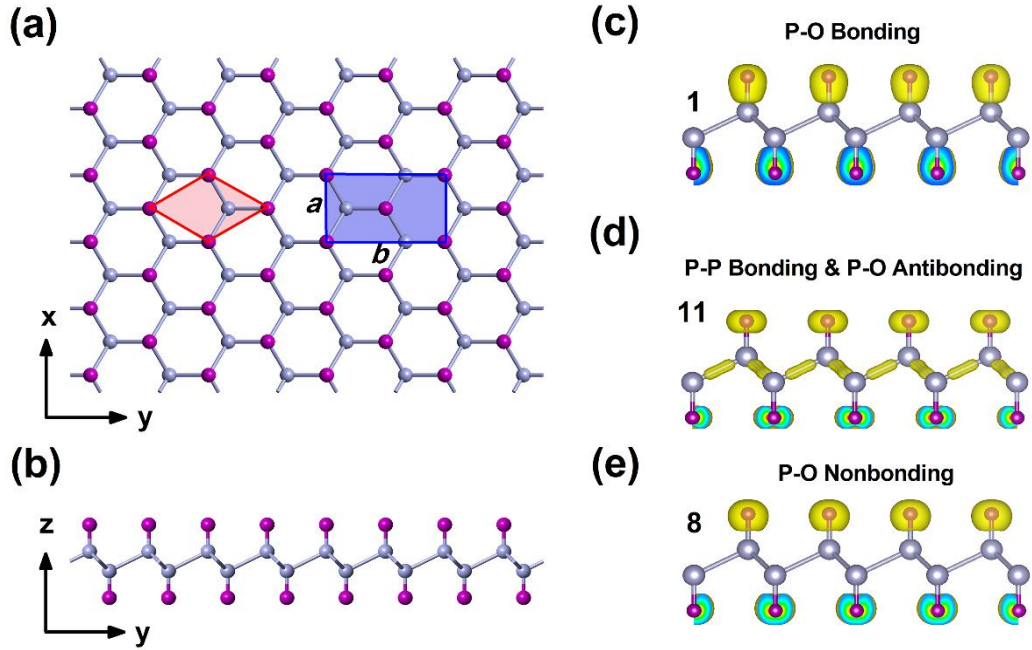

**Fig. 1** (a, b) Top view and side view of the monolayer BPO, where cyan and purple balls represent the P and O atoms respectively. The red region and blue region represent the unit cell and rectangle supercell, respectively. (c, d, e) Three typical charge density distribution. The isovalue of charge density is set as $1.5\times10^{-2}$ eÅ$^{-3}$. The number on the upper left represents the sequence number of band.

Due to the centrosymmetric nature of BPO, one simplified tetrahedron unit with O atom is displayed in Fig. 2(a) and Fig. S2. Fig. S2(a) shows the response of lattice constant to the vertical movement of O atom with fixed z coordinate of P atoms. It can be seen from Fig. S2(b) that the lattice constant of BPO is expanded (contracted) when the P-O bond length is decreased (increased). The strength of P-O bonding state is dominated by the P-O bond length ($L$). We can also say that $L$ is the characteristic length of P-O bonding state; correspondingly, the P-P bond length is the characteristic length of P-P bonding state. For P-O repulsion effect, given that the P-P bonding state brings the electrons together around the midpoint of P-P bond, its characteristic length for P-O ($CL_{P-O}$) repulsion effect is better described as:

$$CL_{P-O} = L+wH \qquad (1)$$

Where $H$ is the vertical height of the P-P bond. Coefficient $w$ is a deliberately introduced parameter, referring to the contribution from $H$ to $CL_{P-O}$. If only the P-O antibonding state is considered, the $w$ is 0. If the electrons are concentrated exactly in the middle of the two P atoms and the $w$ can be estimated as 0.5. When $L$ is compressed and $H$ remains the same (the $z$ coordinate of P atoms is fixed), the characteristic length of P-O repulsion effect decreases, then the repulsion effect is enhanced resulting in a lattice expansion, and vice versa. All these confirm that the P-O repulsion effect exists and has a great influence on the lattice constant.

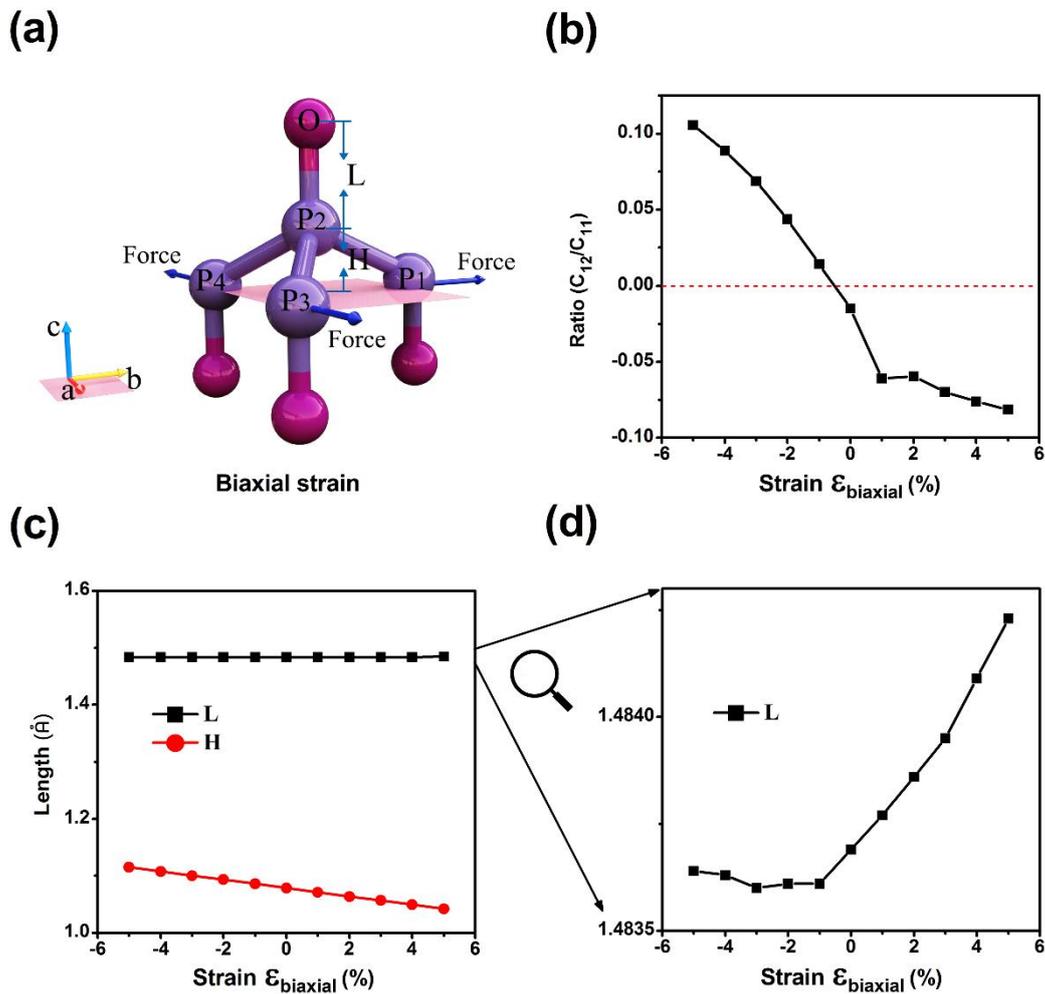

**Fig. 2** (a) Schematic diagram of BPO under biaxial strain. $L$ is the bond length of P-O and $H$ is the vertical height between the P-P. The ratio $C_{12}/C_{11}$ (b), $L$ and $H$ (c) as a function of biaxial strain. (d) Locally enlarged view of $L$.

Starting with the relaxed structures, BPO is deformed with biaxial strain in the range of ±5% to explore its mechanical properties. The strain (biaxial strain, uniaxial strain, or resultant transverse strain) are defined as $\varepsilon = (a_{strain} - a_0)/a_0$, where $a_{strain}$ and $a_0$ is lattice constant with and without strain, respectively. The calculated elastic constant $C_{11} = C_{22} = 148.58$ kbar, $C_{12} = -2.21$ kbar without strain, and the values of elastic constant under all deformation conditions satisfies the mechanical stability criteria $C_{11} = C_{22} > C_{12}$ for the hexagonal case[54] as shown in Tab. 1. It should be noted that, for 2D materials with hexagonal symmetry, the PR can be written using stiffness tensor ($-S_{12}/S_{11}$) or elasticity tensor ($C_{12}/C_{11}$). Since the $C_{ij}$ can be obtained directly via VASP,[39] we use the $C_{ij}$ to calculate the PR. It was found that the $C_{12}/C_{11}$ (equivalent to PR of deformed structure) is positive when $-5\% \leq \varepsilon_{biaxial} \leq -1\%$ and becomes negative when $\varepsilon_{biaxial} > 0$ as shown in Fig. 2(b). As the strain goes from -5% to 5%, the $C_{12}/C_{11}$ decreases a lot, which is largely affected by two factors: the P-P attraction and P-O repulsion effect. While the effect of P-O bonding state on mechanical properties can be ignored because their localized charge distribution is insensitive to the strain. The P-P bonding states attract the P atoms towards each other. As strain increases, on the one hand, the P-P bond length is elongated and the transverse direction would be contracted to conserve bond length and release energy, which causes positive $C_{12}/C_{11}$, On the other hand, the elongated P-P bond length weakens the strength of P-P attraction, resulting in the decreasing of $C_{12}/C_{11}$. The strength of P-O repulsion effect is increased owing to the decreasing of $H$ with increasing of strain. However, it is difficult to quantitatively evaluate this amount of repulsion force between the P and O atom; an indirect way to estimate it is to measure the response of $L$ to strain. Compared to $H$, which reduces rapidly, $L$ remains fairly static as shown in Fig. 2(c). While the enlarged view of $L$ as a function of strain, as shown in Fig. 2(d), presents an interesting phenomenon that $L$ remains basically unchanged when $-5\% \leq \varepsilon_{biaxial} \leq -1\%$, indicating that the repulsion force changes little in this strain range, but increases rapidly under positive strain, indicating that the repulsion force increases rapidly when $\varepsilon_{biaxial} > 0$. This nonmonotonic increment of $L$ may be attributed to the coefficient $w$ is not constant (equation (1)) under strain. The increased repulsion force not only elongates $L$ but also pulls the P-P plane to be more flatten, indicating that the transverse direction would be expanded (negative $C_{12}/C_{11}$). The decreased strength of P-P attraction and increased strength of P-O repulsion effect modulate the $C_{12}/C_{11}$ from positive to negative together under biaxial strain.

**Tab. 1** $C_{11}$, $C_{12}$, and $C_{12}/C_{11}$ of BPO under biaxial strain. The unit of elastic constant is kbar.

| Strain | -5% | -4% | -3% | -2% | -1% | 0% | 1% | 2% | 3% | 4% | 5% |
|---|---|---|---|---|---|---|---|---|---|---|---|
| $C_{11}$ | 247.87 | 226.63 | 205.65 | 185.16 | 165.79 | 148.58 | 129.86 | 119.53 | 109.26 | 100.54 | 92.84 |
| $C_{12}$ | 26.23 | 20.14 | 14.11 | 8.10 | 2.39 | -2.21 | -7.94 | -7.14 | -7.61 | -7.64 | -7.54 |
| $C_{12}/C_{11}$ | 0.11 | 0.09 | 0.07 | 0.04 | 0.01 | -0.01 | -0.06 | -0.06 | -0.07 | -0.08 | -0.08 |

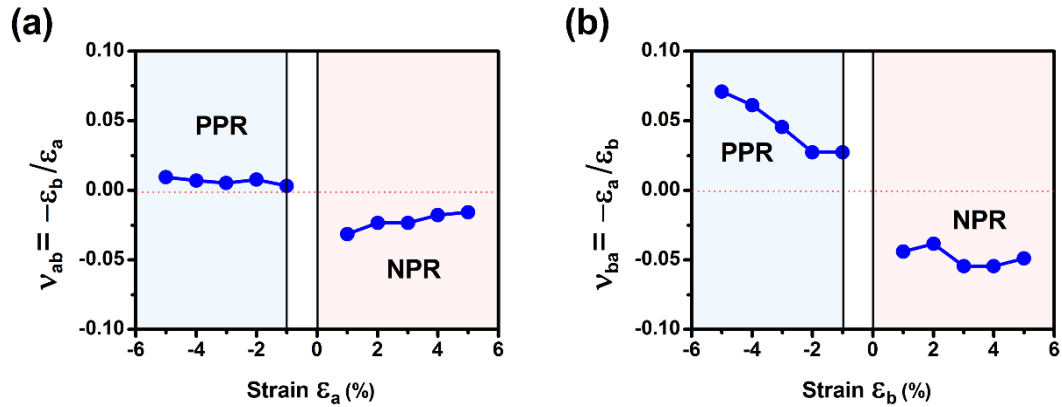

**Fig. 3** PR of BPO as a function of strain applied along *a* direction (a) and *b* direction (b).

We now turn to study the mechanical response under uniaxial strain. The uniaxial strain within the range of $\pm 5\%$ was applied in both the *a* direction and *b* direction, then the lattice constant in the transverse direction was re-optimized. The PR can be obtained as $v_{ab} = -\dfrac{\varepsilon_b}{\varepsilon_a}$ ($v_{ba} = -\dfrac{\varepsilon_a}{\varepsilon_b}$), where $\varepsilon_a (\varepsilon_b)$ is the strain along the *a* (*b*) direction. It was observed that the PR sign of BPO is strongly dependent on the magnitude of applied strain rather than direction. The PR is positive when $-5\% \leq \varepsilon_{uniaxial} \leq -1\%$ and becomes negative when $\varepsilon_{uniaxial} > 0$ as shown in Fig. 3. The structure parameters, including elastic constant and PR, should vary continuously under small strain. Therefore, there should be a sign conversion of the PR when $-1\% < \varepsilon_{uniaxial} < 0$. The variations of $L$ with strain share the same trend with that under biaxial strain (not shown).

It has been reported that the BPO experience band structure transition from semiconductor to semimetal under strain[43]. As an example, we plotted the band structure and density of state (DOS) of BPO supercell under uniaxial strain along $a$ direction, as shown in Fig. S3. We think the electronic phase transition and mechanical properties of BPO are not much related, based on the following two points. On the one hand, the band gap of BPO is closed under uniaxial strain in the range of $1\% \leq \varepsilon_a \leq 2\%$, as shown in Fig. S3(b). The critical biaxial strain for gap closing is bigger (3.3%) on HSE06 level.[43] The strain required for band transition and sign change of PR are not in the same strain range. On the other hand, the band transition indicates the electron transfer in the BPO. As shown in Fig. S3(c), the electron occupying around states B without strain would transfer to other states under strain of 5%. While these transferred states are too tiny, due to small energy range and negligible DOS between the energy of state B and the fermi level (Fig. S3(d)), so that it can be ignored.

The deformation mechanism under uniaxial strain also can be explained by the mutual competition between P-P attraction and P-O repulsion effect. As an example, the deformation evolutions under strain applied in $a$ direction are schematically shown in Fig. 4(a, b). Both the PPR caused by the P-P attraction and NPR caused by P-O repulsion effect under strain can be illustrated by three-step deformation process marked with sequence number, as ①-⑥. It should be noted that the steps for NPR occur simultaneously with that of PPR, but marked with larger sequence number to distinguish them. The PPR induced by the P-P attraction can be understood as the deformation process is to conserve the bond length, there are many similarities with the normal deformation in 1T TMDs.[38] The steps are as follows: ① $P_2$ relaxes along the line $P_2M$ to conserve the bond length of $P_2P_3$ and $P_2P_4$, which are elongated by uniaxial strain along the $a$ direction. Due to the bond angle of $P_1P_2M > 90°$, the $P_1P_2$ is elongated. ② The line $P_2M$ rotates clockwise around the axis of $P_3P_4$ to conserve the bond length of $P_1P_2$. Noting that the location of $P_1$ is fixed in the first two steps. After these two steps, the bond length of $P_1P_2$ still has a certain growth. ③ $P_1$ is contracted accompanied with the reduction of angle $\angle P_1P_2P_3$ to release the storing energy in enlarged $P_1P_2$. The NPR induced by P-O repulsion effect also can be considered as three steps as shown in Fig. 4(b), which are as follows: ④ $P_2$ moves downward, which is the sum of vertical component of steps ① and ②. ⑤ The strength of P-O repulsion effect is greatly increased due to the decreasing of $H$, which pulls the O and P-P plane away from each other. Thus the bond length of P-O is increased and the P-P plane becomes more flatten. ⑥ The angle $\angle P_1P_2M$ and $\angle P_1P_2P_3$ increase and the lattice constant along $b$ direction expands. The deformation evolution under strain can also be reflected in three structural parameters: $H$, $\angle P_1P_2M$, and $\angle P_1P_2P_3$. For $H$ and $\angle P_1P_2M$ (Fig. 4(c)), these two parameters themselves vary greatly with strain, which covers the effect of PR change on it. Therefore, we plot the deviation of these two parameters as shown in Fig. 4(d). As strain increases, the decreasing of $H$ and the increasing of $\angle P_1P_2M$ become much faster under tensile strain (maximum absolute value of the differential value when $-5\% \leq \varepsilon_{uniaxial} \leq -1\%$ is less than minimum absolute value of it when $\varepsilon_{uniaxial} > 0$), in agreement with our expectation that enhancing of P-O repulsion effect pulls the P-P plane to be more flatten. More significant evidence is that the $\angle P_1P_2P_3$ possesses obvious mutations as the strain goes from -1% to 0 as shown in Fig. 4(e), which is consistent with the sign conversion of PR in this strain range.

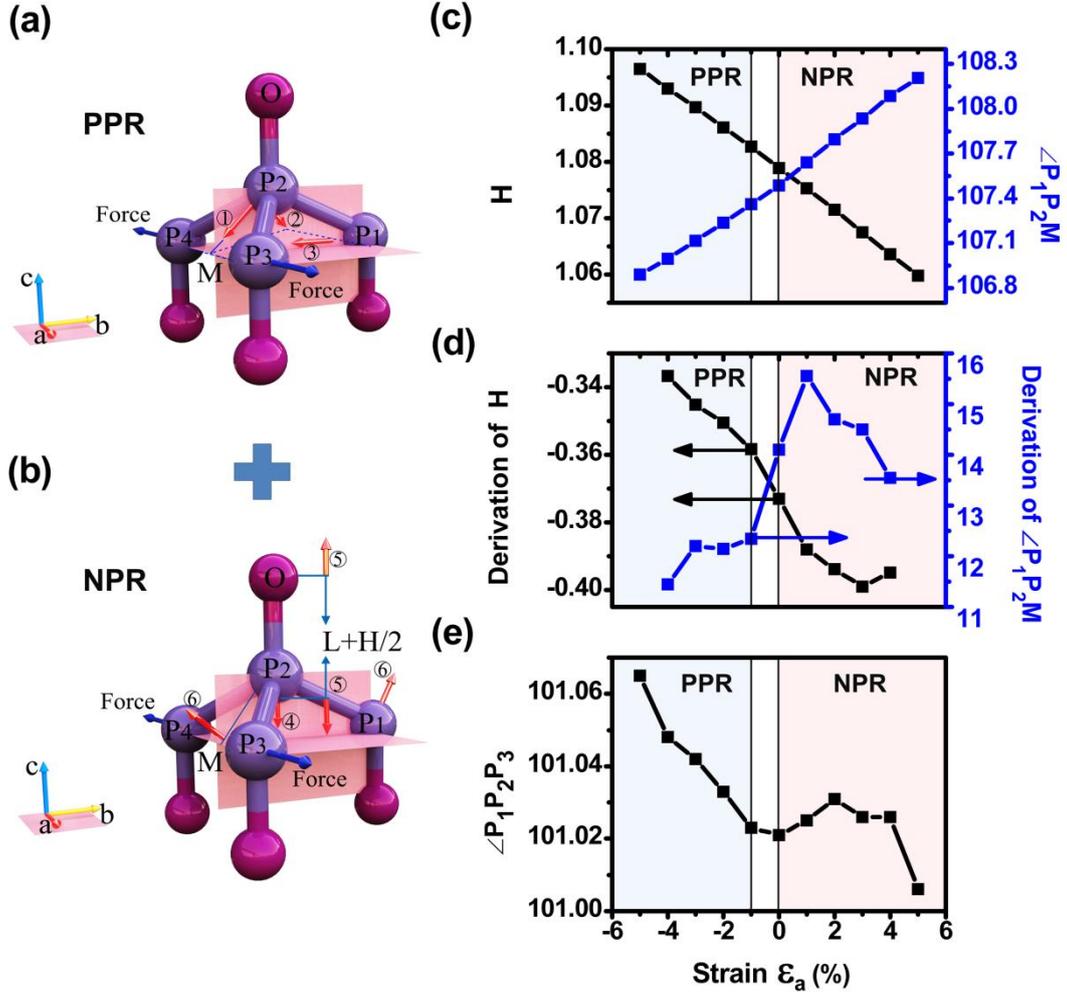

**Fig. 4** Deformation mechanism for PPR (a) induced by P-P attraction and NPR (b) induced by P-O repulsion effect. The blue and red arrow represent the loading strain and deformation evolution, respectively. $H$ and $\angle P_1P_2M$ (c) and their derivations (d) as a function of uniaxial strain. The horizontal arrow represents the maximum or minimum value of differential value in these two strain ranges. (e) $\angle P_1P_2P_3$ as a function of uniaxial strain.

All the absolute value of PR of BPO are smaller than 0.1 under ±5% uniaxial strain, which is much lower than that in some bulk materials. We have listed the average PR (half of the sum of PR in two in-plane directions) of some monolayer NPR materials (we know so far) as shown in Tab. 2. It can been seen that many of these absolute value of PR are ≤ 0.1, which may be attributed to large surface effect in monolayer materials. In the previous study, Ho *et al* have reported that surface effect can strongly influence the lateral strain in metal nanoplates when extended.[25] As the thickness of aluminum nanoplates decreases, which also means that the proportion of surface effect increases, the PR as a function of strain gradually becomes smooth (the value of maximum NPR is decreased).[25]

**Tab. 2** Average PR of some monolayer materials with in-plane NPR.

| | |
|---|---|
| Penta-(graphene, $B_2N_4$, $B_4N_2$) | $-0.068^{55}$, $-0.02^{56}$, $-0.19^{56}$ |
| δ-P | $-0.213^{57}$ |
| $Zn_2C$ | $-0.007^{58}$ |
| Borophenes | $-0.03^{59}$ |
| $Be_5C_2$ | $-0.10^{60}$ |
| Semi-fluorinated graphene | Maximum : $-0.053^{61}$ |
| 1T Mo(W, Te, Re)$S_2$($Se_2$,$Te_2$) | $(-0.03 \sim -0.37)^{38}$ |
| Seven of them | $< -0.1^{38}$ |

The sign change of PR in BPO is similar to the intrinsic deformation pathways from angle stretching mode (with PPR) to bond stretching mode (with NPR) as strain increases in graphene, which is explained by the deformation pathway with NPR with lower energy when $\varepsilon_{uniaxial} > 6\%$ .[62] The difference is that herein the analysis of deformation pathway is based on the electronic effect and the O atom plays an important role through P-O repulsion effect. It should be noted that the PR of BlueP is always positive as shown in Fig. S4. The introduction of O atoms not only changes the electronic structure, but also causes quite different mechanical response under strain. The mechanism we proposed for NPR suggest chemical functionalization combined with strain could also effectively regulate the PR.

**Conclusions**

In conclusion, on the basis of first principle calculations, we found stress-sign-tunable PR, which is positive under $\varepsilon_{uniaxial} <= -1\%$ but becomes negative under $\varepsilon_{uniaxial} > 0$, in monolayer BPO. This sign change of PR is ascribed to mutual competition between the P-P attraction and P-O repulsion effect. As the uniaxial strain goes from -5% to 5%, the P-P attraction causes PPR to conserve the bond length, and its strength is decreased due to the elongated P-P bond length. While the P-O repulsion effect pulls the P-P plane more flatten and leads to NPR, and its strength gradually increases due to the decreasing of *H*. The decreased strength of P-P attraction and increased strength of P-O repulsion effect with increasing of strain modulate the PR of BPO from positive to negative.

## Acknowledgments

This work is supported by the National Natural Science Foundation of China (Grant No. 21673296), the Natural Science Foundation of Hunan Province (Grant No. 2018JJ2481) and the Fundamental Research Funds for the Central Universities of Central South University (2017zzts065).